\documentclass[preprint,12pt]{elsarticle}
\usepackage{tabularx}
\usepackage{lineno,hyperref}
\modulolinenumbers[5]
\usepackage{amsmath,amsfonts}
\usepackage{algorithmic}
\usepackage{algorithm}
\usepackage{array}
\usepackage{xcolor}
\usepackage[caption=false,font=normalsize,labelfont=sf,textfont=sf]{subfig}
\usepackage{textcomp}
\usepackage{stfloats}
\usepackage{soul} 
\usepackage{url}
\usepackage{verbatim}
\usepackage{graphicx}
\usepackage{tabularx,booktabs,textcomp}
\usepackage{amsthm} 
\usepackage{longtable}
\usepackage{tabularray}

\usepackage{amsmath,amssymb,amsthm}

\journal{Elsevier}

\begin{document}

\begin{frontmatter}

\title{Enhanced Rapid Detection of High-impedance Arc Faults in Medium Voltage Electrical Distribution Networks}
\author[1,2]{Kriti Thakur}

\author[1,5]{Divyanshi Dwivedi}

\author[1,2]{K. Victor Sam Moses Babu}

\author[2]{ Alivelu Manga Parimi}
\author[3,4]{Prasanta K. Panigrahi}
\author[5]{Pradeep Kumar Yemula}
\author[2]{Pratyush Chakraborty}
\author[1]{Mayukha Pal\corref{mycorrespondingauthor}}
\ead{mayukha.pal@in.abb.com}
\cortext[mycorrespondingauthor]{Corresponding author}


\affiliation[1]{organization={ABB Ability Innovation Center},
     addressline={Asea Brown Boveri Company},
    city={Hyderabad},
    postcode={500084},
    state={Telangana},
    country={India}.}    

\affiliation[2]{organization={Department of Electrical and Electronics Engineering},
    addressline={Birla Institute of Technology and Science, Pilani- Hyderabad Campus},
    city={Hyderabad},
    postcode={500078},
    state={Telangana},
    country={India}.} 
\affiliation[3]{organization={ Department of Physical Sciences},
     addressline={Indian Institute of Science Education and Research },
    city={Kolkata, Mohanpur, Nadia},
    postcode={741246},
    state={West Bengal},
    country={India}.} 
    \affiliation[4]{organization={ Centre for Quantum Science and Technology},
     addressline={Siksha ‘O’ Anusandhan university},
    city={ Bhubaneswar},
    postcode={751030},
    state={ Odisha},
    country={India}.}

\affiliation[5]{organization={Department of Electrical Engineering},
    addressline={Indian Institute of Technology},
    city={Hyderabad},
    postcode={502205},
    state={Telangana},
    country={India}.}

\begin{abstract}

{ High-impedance arc faults in AC power systems have the potential to lead to catastrophic accidents. However, significant challenges exist in identifying these faults because of the much weaker characteristics and variety when grounded with different surfaces. Previous research has concentrated predominantly on arc fault detection in low-voltage systems, leaving a significant gap in medium-voltage applications. In this work, a novel approach has been developed that enables rapid arc fault detection for medium-voltage distribution lines. In contrast to existing black-box feature-based approaches, the Hankel alternative view of the Koopman (HAVOK) analysis developed from nonlinear dynamics has been applied, which not only offers interpretable features but also opens up new application options in the area of arc fault detection. The method achieves a much faster detection speed in 0.45 ms, 99.36\% enhanced compared to harmonic randomness and waveform distortion method, thus making it suitable for real-time applications. It demonstrates the ability to detect arc faults across various scenarios, including different grounding surfaces and levels of system noise, boosting its practical importance for stakeholders in safety-critical industries.}

\end{abstract}

\end{frontmatter}

\section{Introduction}
\label{section:Introduction}
The International Energy Agency (IEA) reports a worldwide electricity consumption rise of about 6\% during 2021
, with an average yearly growth of 2.7\% projected from 2022 to 2024 \cite{iea2022electricity}. The growing power demand has resulted in a growing number of safety hazards. Therefore, it is crucial to protect the electrical distribution system  \cite{MP_USpatent, MP_WOpatent}. The predominant fault type in the distribution network is the single-phase grounding fault, which constitutes around 80\% of the overall fault occurrences. A high impedance fault (HIF) is a frequently occurring form of single-phase grounding fault. It is characterized by a wire coming into contact with a surface made of a material with high impedance \cite{wang2019high}. The HIF process frequently produces an arc, which is followed by the emission of heat as well as light. Therefore, it is often referred to as a high-impedance arc fault (HIAF) \cite{zeng2023high}. With the integration of multiple energy sources into the power grid, the operating environment becomes more complicated and unpredictable, further increasing the likelihood of HIAF events \cite{hyun2022arc}.

An arcing fault is a particular type of fault that occurs in medium voltage (MV) distribution lines \cite{zhang2016model}. This fault has the ability to create a shock danger, induce equipment failure, and start a catastrophic fire. The occurrence of an arc is directly associated with various types of faults in MV distribution lines \cite{zhang2016model}. It is usually accepted that an incipient fault includes an arc, particularly an incipient fault in an underground cable \cite{kulkarni2014incipient}. The magnitude of the electric current in these arcing faults could reach several kilo-amperes, making them known as high-current arcing faults.
Furthermore, the occurrence of a HIF is consistently associated with an arc, wherein the current of the fault typically ranges from just a few tens of amperes \cite{sedighizadeh2010approaches}. This type of defect is commonly known as a low-current arcing fault. 

Detecting arcing faults in distribution lines is particularly challenging due to their transient nature. To ensure electrical safety, accurate identification of these arc faults is essential. However, in practical engineering applications, arc faults are difficult to detect because of their inherent characteristics, especially their high unpredictability and strong concealment within normal system operations. 

The detection of high-impedance arc faults in MV distribution systems poses significant challenges not typically observed in low-voltage (LV) environments. MV systems, due to their elevated operating voltages and currents, can conceal fault signatures, as these may closely resemble normal load transients. Moreover, high-impedance faults produce fault currents that are significantly reduced due to elevated fault impedance, causing these currents to frequently mimic normal currents, hence making detection challenging. Reliable identification is made more difficult by the short length and variety of these errors, as well as the electromagnetic interference and noise that are common in MV settings. 

Table \ref{tab:arc_fault_overview} highlights the key approaches and constraints discovered in the literature for arc fault detection.As shown in Table \ref{tab:arc_fault_overview}, existing arc fault detection methods face significant drawbacks, including poor generalization, sensitivity to noise and hyperparameters, high computational cost, and limited capacity to capture nonlinear arc dynamics.  In response to these constraints, the proposed HAVOK-based framework utilises Koopman operator theory and delay embedding to extract physically meaningful and dynamically informative features.  By bridging the gap between black-box AI models and conventional physics-based techniques, this methodology delivers better resilience, interpretability, and computational efficiency for high-impedance arc fault detection.

In the past few years, the Koopman operator \cite{korda2018linear} has been utilized in a wide variety of nonlinear time series analysis and has achieved significant success. A notable approach for the investigation of nonlinear dynamical systems is the HAVOK analysis \cite{brunton2017chaos}. HAVOK analysis was proposed by Brunton \cite{brunton2017chaos}. It uses the delay embedding approach and the Koopman theory \cite{article} to reduce chaotic dynamics into a linear model with intermittent forcing.

The HAVOK technique has been widely implemented across numerous fields, displaying its versatility in electrical engineering \cite{dwivedi2023dynamopmu}, and physics \cite{yang2022hybrid}.
 In \cite{dwivedi2023dynamopmu}, the HAVOK method is used to analyze real-time streaming data from electrical distribution networks' distribution-level phasor measurement units. Time series data from behavioral and psychological research\cite{moulder2023extracting} can have substantial nonlinear components, making analysis challenging. To address these difficulties, HAVOK analysis is applied as a flexible framework for capturing and analysing underlying dynamics. By offering a linear framework, recreating attractor manifolds, and estimating the amount of nonlinear forcing in the dynamic system, it makes a more complex analysis possible. These considerations drive the development of increasingly sensitive and interpretable techniques, like the HAVOK-based analysis provided in this work, by requiring sophisticated detection techniques that can spot small, nonlinear irregularities within complicated electrical signals.
 \begin{table}[htbp!]
\centering
\caption{Overview of Arc Fault Detection Strategies and Research Directions}
\begin{tabular}{|p{3.5cm}|p{10.2cm}|}
\hline
\textbf{Research Focus} & \textbf{Description } \\
\hline
System Voltage Level & The majority body of literature concentrates on arc fault detection in low-voltage systems, with limited consideration given to medium-voltage networks \cite{li2023low}. \\
\hline
Detection Strategies & Arc fault detection methodologies are primarily categorised into conventional methods and those employing artificial intelligence (AI) \cite{wang2016high, SOHEILI2018124, wang2022novel}. \\
\hline
Analytical Domains & Detection methods are based on time-domain \cite{wang2016high}, frequency-domain \cite{4150605}, time-frequency domain \cite{ghaderi2014high}, or physical signal features such as sound, light, and electromagnetic radiation. \\
\hline
Traditional Signal Processing Methods & Traditional techniques, including FFT-based harmonic analysis, are utilised to extract arc characteristics.  Nevertheless, these approaches frequently exhibit sensitivity to switching noise and constrained adaptability to fluctuating operating conditions \cite{SOHEILI2018124}.\\
\hline
AI-Based Techniques & Machine learning algorithms, including ANN, SVM, and KNN, have been utilised in conjunction with techniques like as EMD and VMD.  Despite their success, these approaches are usually limited by overfitting, sensitivity to hyperparameters, and the need for large training datasets \cite{lala2020detection, wang2021arc, wei2020distortion, wang2018series}. \\
\hline
Physics-Informed Approaches & Physics-driven approaches, such as time-frequency analysis and sparse representation, have been presented as interpretable alternatives to data-driven models.  However, most fail to adequately capture the nonlinear dynamics inherent in arc behavior \cite{ghaderi2014high, 8566009}. \\
\hline
Deployment & Conventional detection algorithms demonstrate considerable computing complexity and are inappropriate for real-time applications in complicated distribution networks.  It is also unfeasible to install arc fault detectors in every branch, complicating the process of feature extraction from main feeder currents \cite{qu2018arc}. \\
\hline
\end{tabular}
\label{tab:arc_fault_overview}
\end{table}

 To overcome the key constraints in existing high-impedance arc fault (HIAF) detection strategies, particularly for medium-voltage (MV) systems, this work proposes a physics-informed, data-driven method using HAVOK analysis, established in nonlinear dynamical systems theory. The study utilises Koopman-based HAVOK analysis to attain precise and comprehensible arc fault identification.  In contrast to conventional black-box methodologies, our technique offers transparent characteristics through the forcing operator, hence improving explainability.  Moreover, it provides rapid detection times, hence enhancing both precision and comprehension of fault dynamics. The primary novel contributions of this work are:

\begin{enumerate}
    \item Application of HAVOK analysis for arc fault detection in MV distribution networks, enabling a linear interpretation of nonlinear system dynamics —a concept previously unexplored in this domain
    \item Rapid detection capability, achieving arc fault detection within as little as 0.45 milliseconds, which is at least 99.36\% enhanced than existing approaches, considerably enhancing detection speed and system responsiveness.
    \item Robust performance over multiple fault situations, including variable grounding materials (e.g., dry soil, wet cement), system noise levels, and fault current types (low and high current), all verified by simulation using PSCAD.

    \item Discrimination between arcing and non-arcing events utilising a Koopman-invariant forcing operator as a fault signature—offering both accuracy and interpretability, a fundamental drawback in past black-box AI-based techniques.

\end{enumerate}
This article has been arranged as follows. Section \ref{section:arcmodel} discusses the development of the arc model in PSCAD software and generating data for a medium voltage distribution system. Section \ref{section:method} discusses the approach utilized for arc fault detection.Section \ref{section:results} presents the experimental scenarios used to validate the effectiveness of the suggested detection algorithm. Section \ref{section:Comparison With Existing Methods} offers an analysis of our proposed method compared to existing approaches presented in the literature.  Section \ref{section:conclusion} gives the conclusion.

\section{Arc Modelling}
\label{section:arcmodel}
\begin{figure*}[h]
    \centering
    \includegraphics[width=13cm]{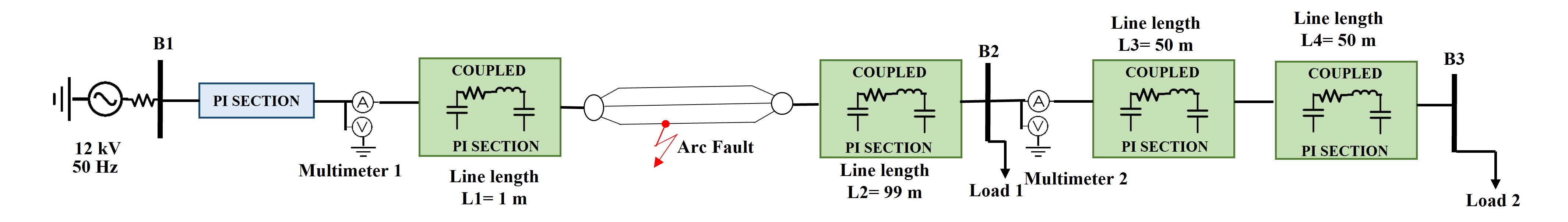}
    \caption{Model for the MV distribution system that includes an arc fault.}
    \label{fig:system}
\end{figure*}

\begin{figure}[h]
    \centering
    \includegraphics[width=6 cm]{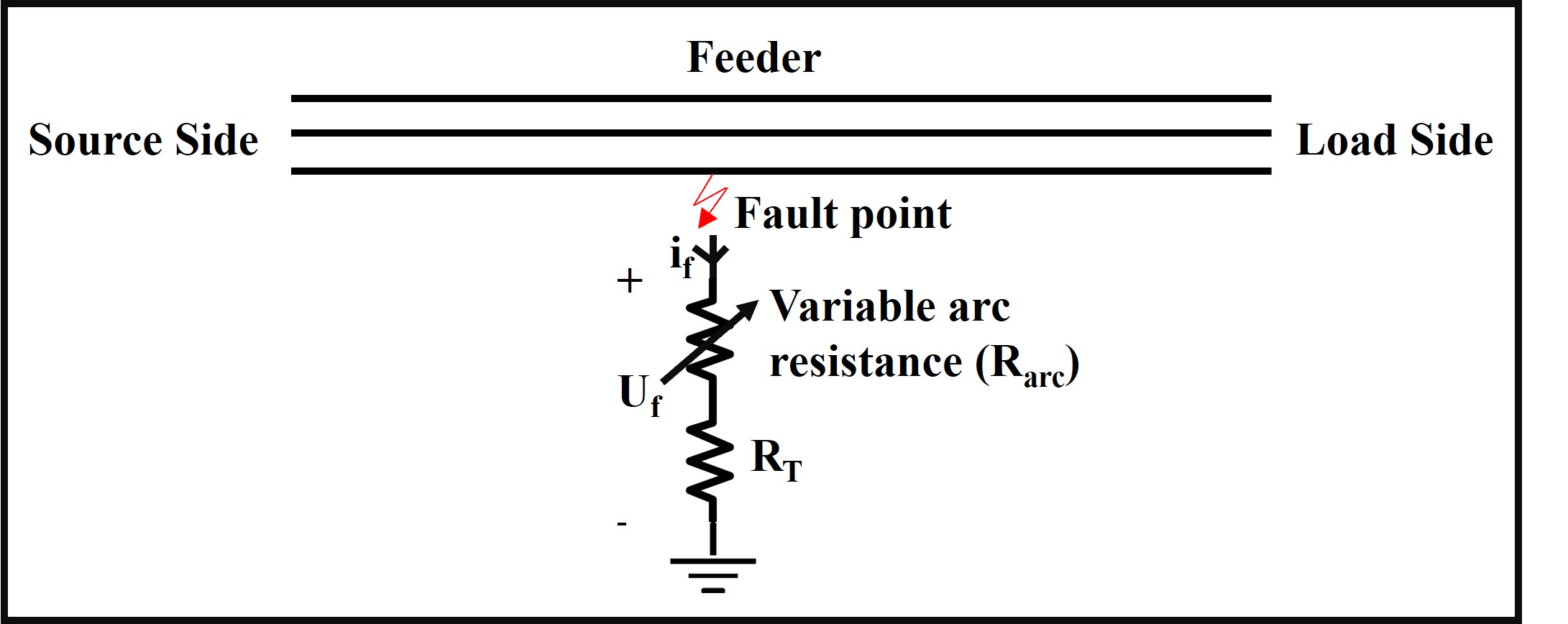}
    \caption{Standard structure of a HIAF model.}
    \label{fig:FIGarc}
\end{figure}

 A Medium-voltage distribution system has been considered that operates at 12 kV and 50 Hz. The simulation was performed in PSCAD software. The system consists of three buses labeled as B1, B2, and B3, and two loads of 10 MW each, shown in Figure \ref{fig:system}. An arc model has been integrated into the system to precisely replicate arc faults. A HIAF model is commonly created by connecting a variable resistor, \(R_{arc}\),  to represent the arc, and a constant resistor to indicate the poor conductivity of the grounding material
, \(R_{T}\) as shown in Figure \ref{fig:FIGarc} \cite{zhang2016model}. Typically, the nonlinear aspects of the arcing process could be represented by \(R_{arc}\), and the current amplitude magnitudes are obtained by modifying \(R_{T}\). The \(R_{T}\) value impacts the amount of current that can flow through the circuit during an arc fault. A greater \(R_{T}\) value often results in reduced current flow, which can influence the detection capabilities of the system. The sensitivity of the detection system to discover arc faults can be modified by the \(R_{T}\) value. In simulations, modifying the \(R_{T}\) helps to explore how varying resistance levels affect the system's response to arc faults. This helps in improving the detection method and analyzing the dynamics of arc faults.

The heat balance equation is typically used to determine \(R_{arc}\) for black-box models \cite{cigre198813}.
\begin{equation}
\label{eq:energy}
     \dfrac{dQ}{dt} =  {u}.{i}- {P}
\end{equation}
where \(P\) (MW) denotes the dissipated power, \(Q\) (MW·s = MJ) denotes the energy that is preserved in the arc
, and \(i\) (kA), and \(u\) (kV) denote the arc current and voltage, respectively. Equation (\ref{eq:energy}) could be subsequently rewritten as:

\begin{equation}
    \dfrac{dQ}{dg} . \dfrac{dg}{dt}  = {P} \bigg(\dfrac{u.i}{P} \ - 1\bigg)
\end{equation}
\begin{equation}
    \dfrac{1}{ g } \dfrac{dg}{dt}  = \dfrac{1}{g.P^{-1}.(\dfrac{dQ}{dg})} \bigg(\dfrac{u.i}{P} \ - 1\bigg)
\end{equation}
where the arc conductance is denoted by \(g\) ( $\Omega ^ {-1}$).
Using the expression ${g.P^{-1}. (\dfrac{dQ}{dg})}$ = $T$, the Mayr model, a widely recognized type of black-box model, could be defined as follows:
\begin{equation}
\label{eq:r}
    \dfrac{lng}{dt}  = \dfrac{1}{T} \bigg(\dfrac{u.i}{P} \ - 1\bigg) =  \dfrac{1}{T'} ({u}.{i}- {P})
\end{equation}
where $T$ is expressed in unit of seconds (s) and $ T'$= $T$ · \(P\).
Several formulas can be used to define the arc parameters $T'$ $T$  and \(P\) depending on the arcing process hypothesis that is used. It has been proven that there is more than one relationship between the arc parameter specifications and the arc characteristics that are presented \cite{schavemaker2000improved}.
The fundamental source of current distortion is the nonlinearity of \(R_{arc}\), the arc resistance. This nonlinearity directly influences the system's behavior and can be expressed mathematically as given in \ref{eq:r}:
\begin{equation}
     \label{eq:arc}
    R_{arc}  = \dfrac{1}{ g } =  e^{\int\limits \dfrac{1}{T'} ({u}.{i}- {P}) \ dt} \
\end{equation}
Therefore $T'$  and the difference between \(P\) and ${u}.{i}$ have an impact on \(R_{arc}\). This difference is known as residual power, and it represents the energy-storing capacity of an arc. 

\begin{figure*}[h]
    \centering
    \includegraphics[width=13cm]{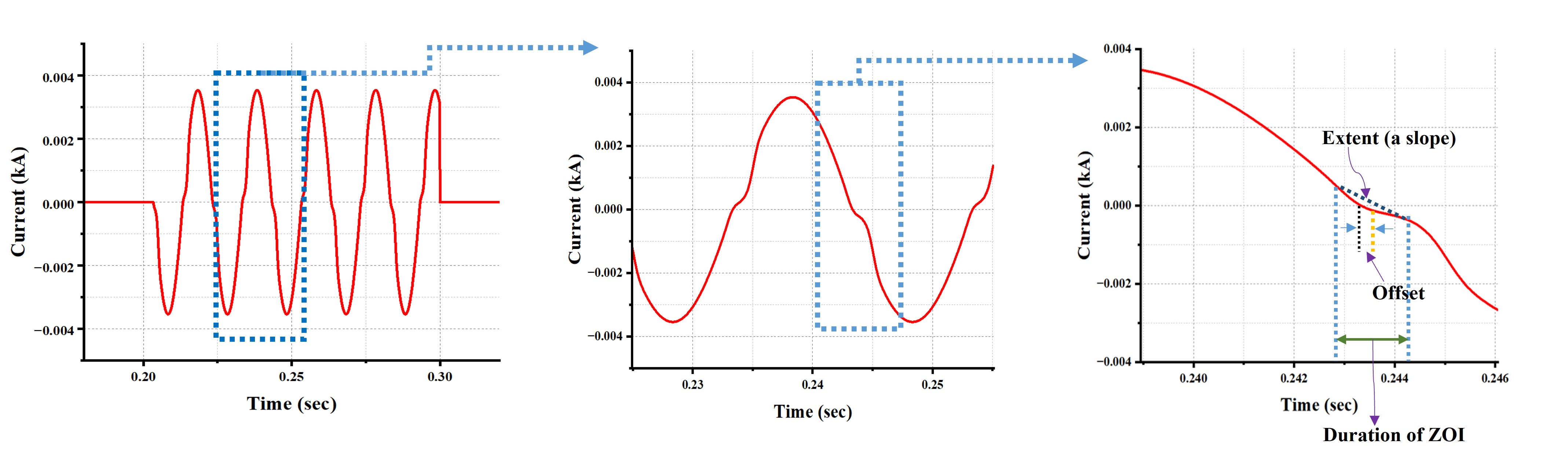}
    \caption{The waveforms of arc current observed in the 12 kV distribution system field.}
    \label{fig:FIG1}
\end{figure*}

The nonlinear nature of arc faults could be changed by three fundamental arc characteristics, each providing a distinct physical interpretation and a means for effectively modifying arc resistance. These characteristics play a key role in determining and characterizing residual power. Consequently, they enable the regulation of three components of current distortions, hence boosting simulation accuracy. Ideally, an ideal current distortion should be nearly centered around the zero point and exhibit symmetry. However, in practical settings, distortion offsets occur relative to higher or lower zero points due to variations in the lag between the dissipated power of the arcs and the energizing power. Consequently, the three primary variables in the current waveforms, as illustrated in Figure \ref{fig:FIG1}, are the duration, extent, and offset of distortions.
Figure \ref{fig:FIG1} includes additional labels to visually depict the three features more understandably.
\begin{enumerate}
    \item DURATION: The term ``duration" describes the period of time of the zero-off interval (ZOI), which is a component of the total distorted interval (DI), allowing models to precisely reflect how long the arc resistance remains at a crucial value, which affects the current waveform. The ZOI waveform aligns with the x-axis when the arc resistance reaches a particular value. 
    \item EXTENT: The slope of the line between the ZOI's beginning and finishing points is known as the extent. It indicates the rate of change of current during the fault event, which is essential for distinguishing arc faults from other fault types.

    \item OFFSET: The offset is the separation that exists between the zero-crossing point and the midpoint of DI. It indicates any shifts in the waveform because of the arc fault.

\end{enumerate}
By effectively managing these characteristics, the simulations produce waveforms that closely resemble actual system behavior, leading to improved detection capabilities, a deeper understanding of fault dynamics, and validation of detection methods. 

Since dissipated power \(P\) has a direct relationship with energizing power, which is determined by ${u}.{i}$ \cite{guardado2005improved}, a mathematical model of \(P\) can be constructed as follows:
  \begin{equation}
    P  = F({ui}) = {u}.{i} + P_{res}(t)
\end{equation}
$P_{res}(t)$ is the residual power function in this case, after which arc resistance \(R_{arc}\) in (\ref{eq:arc}) could be formulated as follows.
\begin{equation}
    R_{arc} =  e^{\int \dfrac{1}{T'} P_{res}(t) \ dt} \
\end{equation}
The residual power could be regulated through the utilization of three current features: Duration, Extent, and Offset. This equation facilitates the adjustment of \(R_{arc}\) throughout the entire simulation period, thereby enabling the generation of waveforms depicting current with an arc fault.
Suppose \(t_m\) denotes the time at which \(R_{arc}\) initially achieves its maximum value in a half cycle; at that time, the residual power is indicated as \(P_{res}(t_m)\).
{According to the characteristics demonstrated by the standard models , $P_{\text{res}}^{(t_m)}(t)$ could be expressed as a periodically segmented linear function:}
\begin{equation}
P_{\text{res}}^{(t_m)}(t) =
\begin{cases} 
a_1 \left( t - t_m - \frac{\text{DURATION}}{2} \right) - b_1, & 
t \in \left[ t_m - \frac{\text{DURATION}}{2},\ t_m + \frac{\text{DURATION}}{2} \right) \\[10pt]
a_2 \left( t - t_m - \frac{\text{DURATION}}{2} \right) - b_2, & 
t \in \left[ t_m + \frac{\text{DURATION}}{2},\ t_m + \frac{\text{Time}}{2} - \frac{\text{DURATION}}{2} \right)
\end{cases}
\label{eq:pres}
\end{equation}

{
where
\begin{equation}
\begin{aligned}
a_1 &= \frac{-8 \ln \text{EXTENT}}{\text{DURATION}^2}, & b_1 &= \frac{4 \ln \text{EXTENT}}{\text{DURATION}}, \\[8pt]
a_2 &= m \cdot \frac{-16 \ln \text{EXTENT}}{\text{DURATION}(Time - 2\text{DURATION})}, & b_2 &= m \cdot b_1
\end{aligned}
\end{equation}
$m$ is a coefficient and $Time$ indicates the amount of time of an interval (20 ms in a 50 Hz system).
}

{
$\text{OFFSET}$ is used to control the distortion offset. The moment at which the maximum arc resistance occurs $t_m$ is controlled by $\text{OFFSET}$ and satisfies:}

{
\begin{equation}
\begin{cases}
|u_f(t_m)| = \text{OFFSET}, \\[5pt]
\text{sgn}(\text{OFFSET}) \cdot \frac{d|u_f(t_m)|}{dt} \geq 0
\end{cases}
\end{equation}
}

{
Where $u_f$ (kV) indicates the voltage of the fault.
EXTENT: Distortion extent is determined by the amount of the maximum resistance, i.e., $R_{\text{arc}}(t_m)$.  It is calculated as:
}

{
\begin{equation}
R_{\text{arc}}(t_m) = R_{\text{arc},0} + e^{\int_{t_0}^{t_m} P_{\text{res}}(t) \, dt}
\label{eq:k}
\end{equation}
}

{
where, $R_{\text{arc},0} = R_{\text{arc}}(t_0)$, and indicates the minimal resistance during $[t_m - Time/2, t_m]$. $R_{\text{arc},0}$ is the resistance when an arc is highly conductive, and which could be ignored in comparison to the huge grounding resistance $R_T$ .Thus, put $R_{\text{arc},0} = 1$ and \ref{eq:k} represent as:
}

\begin{align}
\ln R_{\text{arc}}(t_m) 
&= \int_{t_0}^{t_m - \frac{\text{DURATION}}{2}} P_{\text{res}}(t) \, dt 
+ \int_{t_m - \frac{\text{DURATION}}{2}}^{t_m} P_{\text{res}}(t) \, dt \nonumber \\
&= \left( \frac{\text{Time}}{8} - \frac{\text{DURATION}}{4} \right) \cdot m \cdot b_1 
+ \frac{1}{2} \cdot \frac{\text{DURATION}}{2} \cdot b_1 \nonumber \\
&= \left( \frac{\text{Time}}{8} - \frac{\text{DURATION}}{4} \right) \cdot m \cdot b_1 
+ \ln(\text{EXTENT})
\label{eq:lnRarc}
\end{align}

{
Hence, from the equation above, the maximum resistance that arises at $t_m$ can be individually reflected by $\text{EXTENT}$ if $m$ is taken as 0.
}

\begin{figure}[h]
    \centering
    \includegraphics[width=8cm]{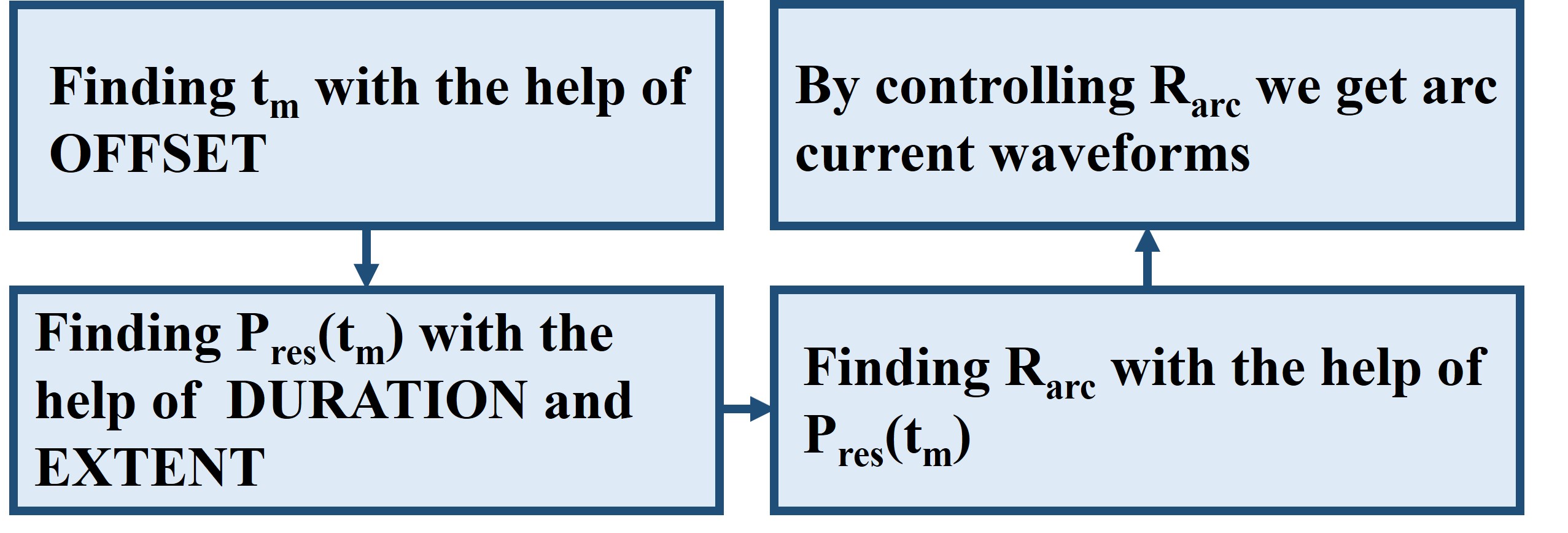}
    \caption{Modelling of an arc fault.}
    \label{fig:arccc}
\end{figure}

{
DURATION: Distortion duration is determined by the length of ZOI, which is exactly equal to $\text{DURATION}$.} {Hence, the distorted duration also becomes individually controlled 
As a result, independent control of all three arc characteristics theoretically make it more flexible in controlling distortion characteristics and achieving the different waveforms.}{In summary, the modelling of an arc fault entails a systematic process starting with the identification of the time instance \( t_m \) via the OFFSET parameter, which indicates the initial appearance of arc fault characteristics. Utilising this \( t_m \), the residual power \( P_{\text{res}}(t_m) \) is computed based on DURATION and EXTENT, capturing the temporal and magnitude aspect of the arc fault. The residual power is utilised to calculate the arc resistance \( R_{\text{arc}} \), so establishing a correlation between power dissipation and resistance during the fault. Ultimately, by regulating \( R_{\text{arc}} \), arc current waveforms are generated, enabling a realistic simulation of arc faults for analysis and validation.}
The arc modelling is given in Figure \ref{fig:arccc}.
\section{Method}
\label{section:method}
The suggested approach is well-suited for examining the dynamic response and showing the nonlinear properties inherent in the electrical distribution system. To illustrate this, take a continuous-time dynamical system with a state vector \(y(t)\),

\begin{equation}
\label{eq:1}
\frac{d}{dt} y(t) = f(y(t))
\end{equation}

The variable $y(t) \in \mathbb{R}^n$ reflects the state of the system at time $t$. The function $f$ describes the dynamics of the state variable $y$. Equation \ref{eq:1} depicts a continuous-time dynamical system evolving over time. The updated formula for the discrete-time acquisition of state samples from current data is written as \cite{eeg}.

\begin{equation}
y_{k+1}=F(y_k)= y_k + \int_{k \Delta t}^{(k+1) \Delta t}  f(y(\tau))d\tau
\label{eq:k}
\end{equation}

In the provided expression, the term \(y_k = y(k \Delta t)\) indicates the sampled point of the system's trajectory generated from equation (\ref{eq:k}), where \(k\) is an integer index and \(\Delta t\) signifies the time interval between consecutive samples. The symbol \(F\) denotes a discrete-time propagator. The variable \(\tau \) denotes the integration variable in the integral representation of the system dynamics. This describes the evolution of the system state during a discrete time step \(\Delta t\), incorporating the impacts of the function $f(y(\tau))$, which characterizes the system's dynamics during that period.

{

\subsection{Arc Fault Detection using Hankel Alternative View of Koopman Analysis}

The delay embedding approach is utilized to reconstruct the state space of a dynamical system from a time series. The basic concept is to produce a multi-dimensional representation of the time series data, which represents the temporal dependencies and dynamics of the system. In our work of arc fault detection, the time-series current data \(i(t)\) taken from system Figure \ref{fig:system}, the time shifted measurement of \(i(t)\) is utilized and stack them to generate the Hankel matrix \(\mathbb{H}_m\), as explained in \cite{koopman}.}In our approach, the Hankel matrix is formed by aggregating 40 time-shifted rows of the input current signal, so establishing the embedding dimension (q) at 40.  This value provides an effective compromise between adequately recording temporal dynamics and ensuring computational efficiency, as informed by previous HAVOK investigations \cite{havok,koopman}. An embedding dimension of 40 was selected based on earlier HAVOK studies to represent the system's dynamic behavior well.  In practice, this number should be tuned to the sample rate.  A greater sampling rate provides for a larger embedding dimension to resolve finer dynamics, whereas a lower rate may require fewer delays.  To guarantee significant reconstruction of the system's attractor, the total embedding window (embedding dimension × sampling interval) should, as a general rule, span at least one cycle of the main system frequency (20 ms for a 50 Hz system, for example). The embedding dimension q=40 was selected based on a trade-off between capturing critical system dynamics and avoiding excessive model complexity.  Similarly, the window length was empirically selected to achieve robust detection performance without representing misleading or non-generalizable fluctuations in the signal\cite{havok}. The window length (p) is dynamically established as the residual signal length following delay embedding, optimizing data usage while maintaining matrix stability.  These parameters facilitate a robust extraction of Koopman-invariant coordinates while preventing overfitting and minimizing excessive noise. Subsequently, eigen time delay coordinates are acquired using the method of singular value decomposition (SVD) applied to the Hankel matrix \(\mathbb{H}_m\).

 \begin{equation}
\mathbb{H}_m= \begin{bmatrix}
i(t_1) & i(t_2) & \dots & i(t_p) \\ 
i(t_2) & i(t_3) & \dots & i(t_{p+1}) \\ 
\vdots & \vdots & \ddots & \vdots \\ 
i(t_q) & i(t_{q+1}) & \dots & i(t_T) \\ 
\end{bmatrix} 
\end{equation}

In the preceding expression, the time points \(t_{j+1}= t_j+\tau\) (where \(j=1, \dots, T-1\)), \(q\) signifies the number of points in the trajectory, and \(p\) represents the window length, representing the largest periodicity obtained by the Hankel matrix. 

Following the application of SVD, the SVD of a matrix \(\mathbb{H}_m\) is the process of decomposing \(\mathbb{H}_m\) into the multiplication of three matrices \( U \Sigma V^T \), where the columns of \( U \) and \( V \) are orthogonal and normalized, and \( \Sigma \) is a diagonal matrix with positive real values. The resulting matrix is the product of these three components.

\begin{equation}
    \mathbb{H}_m= U \Sigma V^T
    \label{eq: 4}
\end{equation}

where $U \in \mathbb {R} ^ {nxn}$, $\Sigma \in \mathbb {R}^{nxm}$, $V \in \mathbb {R} ^ {mxm}$.

The SVD arranges the columns of \( U \) and \( V \) in a hierarchical order based on their ability to best represent the columns and rows of the matrix {$\mathbb{H}_m$} , respectively.
 In general, $\mathbb{H}_m$ gains a low-rank approximation for the first $r$ columns of $U$ and $V$, which is invariant to the Koopman variant for the states.{ Eigen time series in U provides a Koopman invariant measurement system on the attractor. Koopman theory presents a structure for evaluating non-linear dynamical systems using linear approaches. In the proposed methodology, the operation of SVD on the Hankel matrix $\mathbb{H}_m$ reveals that all columns of $\mathbb{H}_m$ are approximately numerically dependent on the leading r columns of U, signifying the presence of a measurement subspace. Thus, the eigen time-delay coordinates in U constitute this measurement subspace.}

{When the Koopman operator acts on any of these measurement coordinates, the resultant representation remains within the same subspace, as all columns of} $\mathbb{H}_m$ {are spanned by the columns of U. This insight results in the development of a data-driven Koopman invariant measuring system, expressed using time-delay coordinates obtained from the Hankel matrix.} $\mathbb{H}_m$ {could be expressed by iterates of the Koopman operator as } $i(t_2) = \mathbb{K}i(t_1)$. {Koopman operator operates on all measurements of our states. So, every element of the Hankel matrix can be described in terms of the Koopman operator applied many times to the initial condition $i(t_1)$. Therefore, we could alternatively rewrite the Hankel matrix as \cite{koopman}:

\begin{equation}
\mathbb{H}_m= \begin{bmatrix}
i(t_1) & \mathbb{K}i(t_1) & \dots & \mathbb{K}^{p-1}i(t_1) \\ 
\mathbb{K}i(t_1) & \mathbb{K}^2i(t_1) & \dots & \mathbb{K}^pi(t_1) \\ 
\vdots & \vdots & \ddots & \vdots \\ 
\mathbb{K}^{q-1}i(t_1) & \mathbb{K}^qi(t_{1}) & \dots & \mathbb{K}^{T-1}i(t_1) \\ 
\end{bmatrix} 
\label{eq:5}
\end{equation}
The columns of equation \ref{eq: 4}, and hence equation \ref{eq:5}, are efficiently approximated by the first \( r \) columns of \( U \), making these eigen-time series a Koopman-invariant measurement system. At the same time, the first \( r \) columns of \( V \) constitute a time series that captures the magnitude of each column of \( U\Sigma \) within the data. A plot of matrix $V$'s first three columns yields an embedded attractor. Then we identify the optimum singular value hard threshold coefficient and compute r. Then the time history of V components is taken, and a regression model on those eigen time delay coordinates. A linear time series that substantially mirrors the input time series is produced by the initial $(r-1)$ variables of $V$ {that gives a very nice regression fit. Rather than developing a closed linear model for the initial r variables in V, a linear model is established based on the first $(r-1)$ variables, and the last variable  $v_r$ gives a bad regression fit, which will function as a forcing term.
\begin{equation}
\frac{d}{dt} \mathbf{v}(t) = \mathbf{A} \mathbf{v}(t) + \mathbf{B} \mathbf{v}_r(t).
\label{eq:6}
\end{equation}
Here $\mathbf{v} = \begin{bmatrix} v_1 & v_2 & \cdots & v_{r-1} \end{bmatrix}^T$ {is a vector of the first $r-1$ eigen-time-delay coordinates. $v_r$ serves as an input forcing that influences the linear dynamics in \ref{eq:6}, resulting in the nonlinear dynamics in \ref{eq:1}. Whenever a nonlinearity exists in the system, it undergoes a bursting phenomenon, which serves as an intermittent forcing signature that is a forcing operator. The matrix A defines the linear dynamics of the system. It defines the evolution of v(t) without external influences. The matrix B serves as a connection between v(t) and the external force $v_r$(t). It describes how the external input (forcing) impacts the evolution of the state.}

The forcing operator captures the current fluctuations during arc faults. Upon initiating an arc, it produces unique electrical signatures characterized by abrupt fluctuations in current intensity and high-frequency noise. The presence of an arc fault is indicated by the emergence of these intermittent forcing bursts, detected using the Forcing Operator. The operator enables the detection of these deviations from the norm. Finally, the forcing operator reflecting the nonlinear characteristics of the arc fault in the current data is provided by the $r^{th}$ column. The forcing operator is the last eigentime delay coordinate that is the $r^{th}$  column produced from the SVD of the Hankel matrix, representing the external driving force influencing the system.  It captures external influences on the system, such as faults or disturbances, which are different from the natural dynamics of the system. By setting a threshold on the forcing operator, arc faults could be efficiently detected.
 Figure \ref{fig:method} provides an overview of the framework proposed in this study. The entire detection procedure is encapsulated in Algorithm 1, which emphasises the formulation of the Hankel matrix, the extraction of forcing coordinates, and the classification of arc faults predicated on the magnitude of the forcing operator.

\begin{figure*}[htp]
  \centering
  \includegraphics[width=5in]{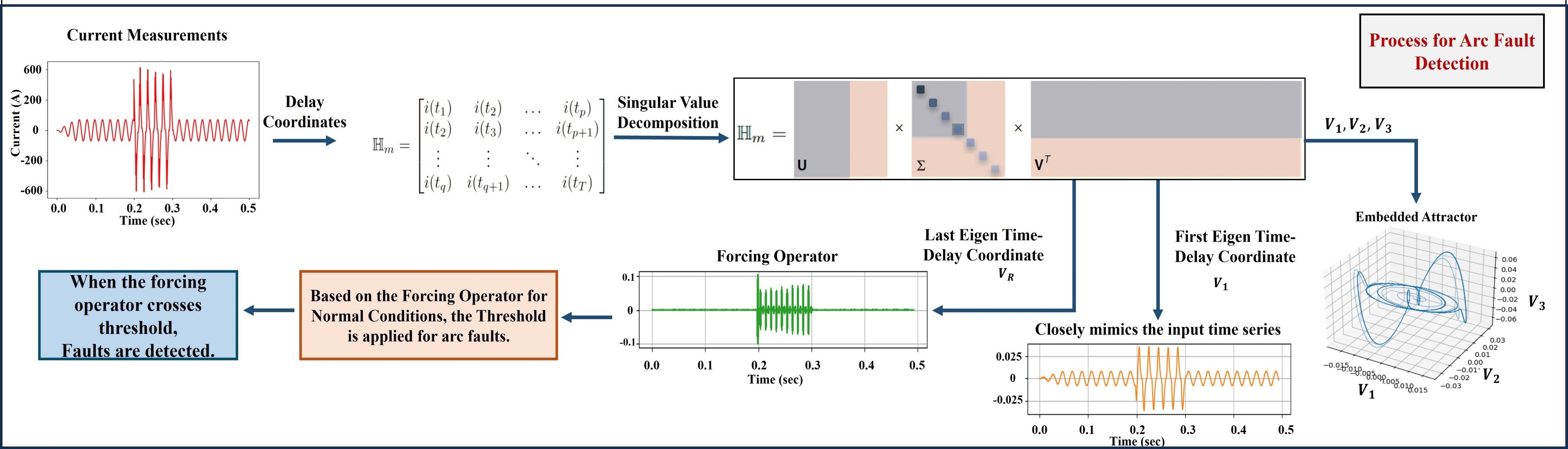}\\
  \caption{An overview of the suggested arc fault detection technique; HAVOK analysis framework is utilized for arc fault detection in current measurements. Time-shifted current measurement is stacked into a Hankel matrix. Its singular value decomposition provides a hierarchy of eigen time series that collectively form a delay-embedded attractor. A best-fit linear regression model is then generated for the delay coordinates \(V\); the linear fit is excellent for the first \(r-1\) variables, however, the last coordinate \(V_r\) is not adequately captured by a linear model. Instead, \(V_r(t)\) functions as a stochastic forcing input to the first \(r-1\) delay coordinates, with few yet significant forcing events corresponding to sudden fluctuations in dynamics that indicate the onset of an arc fault, similar to lobe switching in chaotic systems.}
  \label{fig:method}
\end{figure*}
\begin{algorithm}
\caption{HAVOK-based Arc Fault Detection with Forcing Operator}
\begin{algorithmic}[1]
\REQUIRE Time-series data $i(t)$, sampling interval $\Delta t$, number of delay embeddings $stack\_max$
\ENSURE Detection of arc faults via forcing operator threshold

\STATE Construct the Hankel matrix $H$ by stacking $stack\_max$ delayed versions of $i(t)$
\STATE Perform Singular Value Decomposition: $H = U \Sigma V^T$
\STATE Determine model rank $r$ using an optimal hard threshold on singular values
\STATE Compute derivatives of $V$ using fourth-order central difference method
\STATE Define delay-embedded state matrix $x = V[:, 1:r]$
\STATE Identify the last eigen time-delay coordinate $v_r = x[:, r-1]$ as the forcing operator
\STATE Detect arc fault onset by identifying bursts in $v_r$ beyond predefined thresholds:
\IF{$0.06 \leq |v_r| \leq 0.18$}
    \STATE \textbf{return} Arc fault detected
\ELSIF{$|v_r| > 0.2$}
    \STATE \textbf{return} Other fault type (non-arc)
\ELSIF{$|v_r| < 0.045$}
    \STATE \textbf{return} Non-arcing disturbance (e.g., load switching)
\ELSE
    \STATE \textbf{return} Inconclusive or borderline case
\ENDIF
\end{algorithmic}
\end{algorithm}

Arc faults introduce nonlinear dynamics into the electrical system, which can be tough to detect using typical linear methods. The forcing operator helps to capture these nonlinear interactions by modeling how the system reacts to external influence (the arc fault). The presence of an arc fault can be considered as an intermittent driving force that changes the system's dynamics. By studying the system's response to these driving elements, the HAVOK approach can discover patterns that indicate the presence of arc faults.

\section{ Results and Discussion}
\label{section:results}

The following section provides the findings from the cases conducted to evaluate the effectiveness of the arc fault detection framework. The analysis employs the system illustrated in \ref{fig:system}. The simulation integrates an arc model derived from  \cite{wei2020distortion}, which serves as a fundamental reference for the construction and use of the model in the study. In our study, arc faults were strategically placed at different locations between Bus 1 and Bus 2, as well as between Bus 2 and Bus 3, within the medium voltage distribution system. This placement was done to examine the detection performance across different regions of the network. Altering the fault locations offered an evaluation of the robustness and precision of the arc fault detection system in recognising faults at different points along the distribution line. This approach ensures a full evaluation of the detection system over a range of possible fault conditions. Simulations were carried out at a sample rate of 20 kHz (i.e., 50 µs resolution) to ensure exact time-domain analysis. The algorithm was executed in Python on a normal workstation equipped with an Intel i5 CPU at 2.60 GHz and 16 GB of RAM.

All the studies have included data containing both arc fault and normal current waveforms. As demonstrated in the arc model, variations in the model's variable aspects, particularly DURATION, EXTENT, and OFFSET, result in unique waveforms. The values were set based on an existing reference for a 50 Hz system with a phase-voltage level of 12 kV (the maximum voltage) for our system. The parameter ranges are as follows: 
$\text{OFFSET} \in (-12, 12) \, \text{kV}, \, \text{EXTENT} \in (0, +\infty) \, \text{k}\Omega, \, \text{DURATION} \in (0, 20) \, \text{ms}$
, which aligns with typical fault current behavior in medium voltage systems \cite{wei2020distortion}. The selected values guarantee that the arc fault current graph aligns accurately with the graphs presented in \cite{wei2020distortion}, with the specific values utilised in each case detailed in Table \ref{tab:compare_dataset} for reference. }

The time series data of the current is obtained from PSCAD simulations. The HAVOK analysis is utilized to calculate eigen time delay coordinates from the time series data. The time-delay coordinates are utilized to transform the time-series measurements into the Hankel matrix. Next, singular value decomposition was carried out to extract the columns of U and V in a hierarchical way based on their capacity to represent the columns and rows of the Hankel matrix. Subsequently, the V values were employed to formulate the linear model and forcing operator for unusual occurrences.
The forcing operator facilitates comprehension of how external inputs impact the behavior of the system. Through the identification and investigation of the forcing operator, useful insights have been gained into how various inputs or perturbations affect the observed behavior. These insights are then utilised by applying the range of the forcing operator to detect arc faults using this method. Numerous experiments were conducted, and a threshold was applied to differentiate between various fault types. For arc faults, the range of the forcing operator was measured as +0.06 to + 0.18 and -0.06 to -0.18. In the case of non-arcing disturbances, the range is +0.045 to -0.045. For other types of faults, the forcing operator exhibited a much broader range of +0.2 and higher values and -0.2 and lower values.}
 The simulation runs for 0.5 seconds, with the fault imposed at 0.2 seconds, and remains for 0.1 seconds. The different simulation circumstances are taken into account for analysis.

\subsection{Case A: Low Current Arcing Fault Present} 
\begin{figure*}[htp]
  \centering
  \begin{minipage}[b]{0.45\textwidth}
    \includegraphics[width=\textwidth]{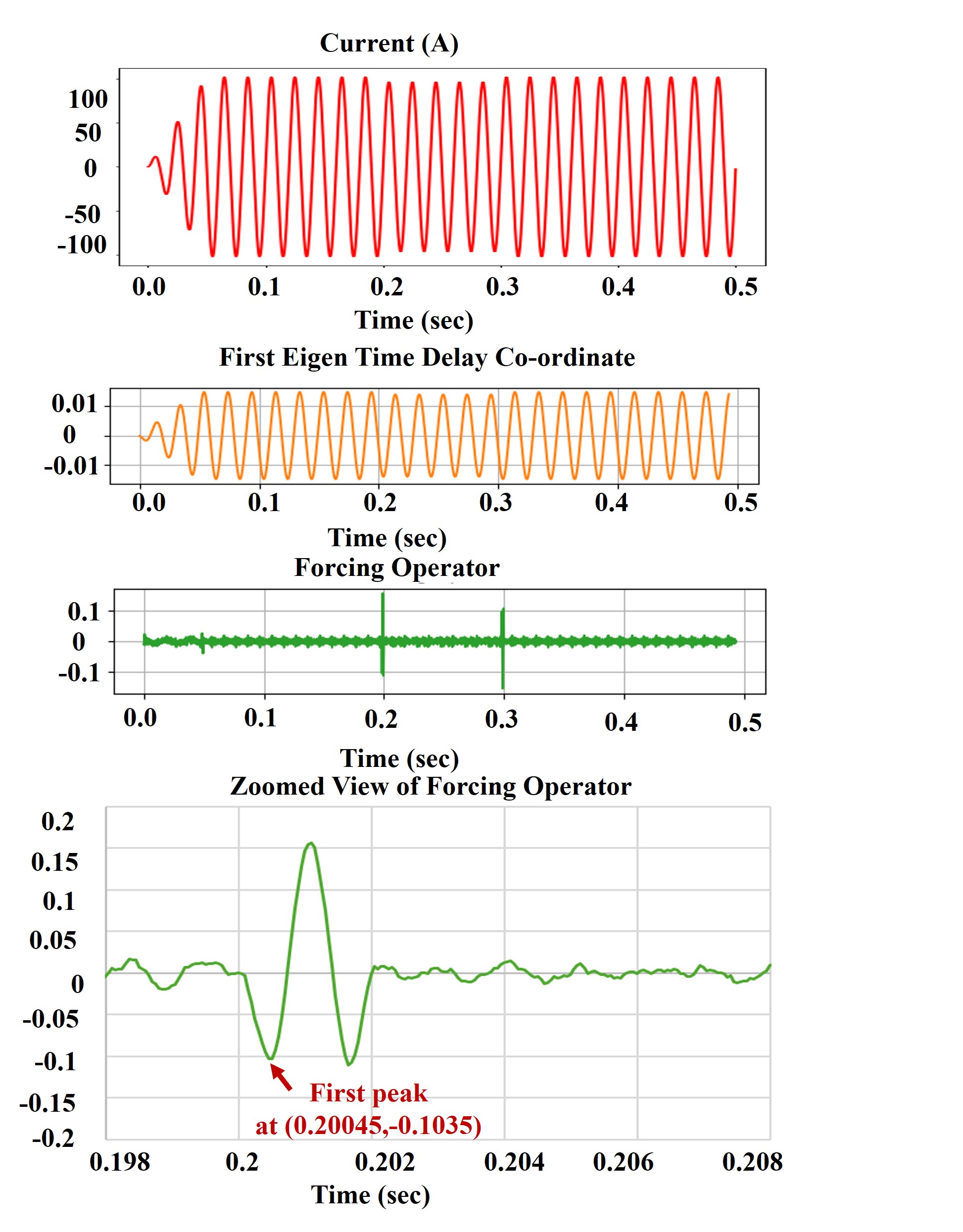}
    \caption{Low current Arcing fault with $R_T = 1000\,\Omega$.}
    \label{fig:FIG2}
  \end{minipage}
  \hfill
  \begin{minipage}[b]{0.45\textwidth}
    \includegraphics[width=\textwidth]{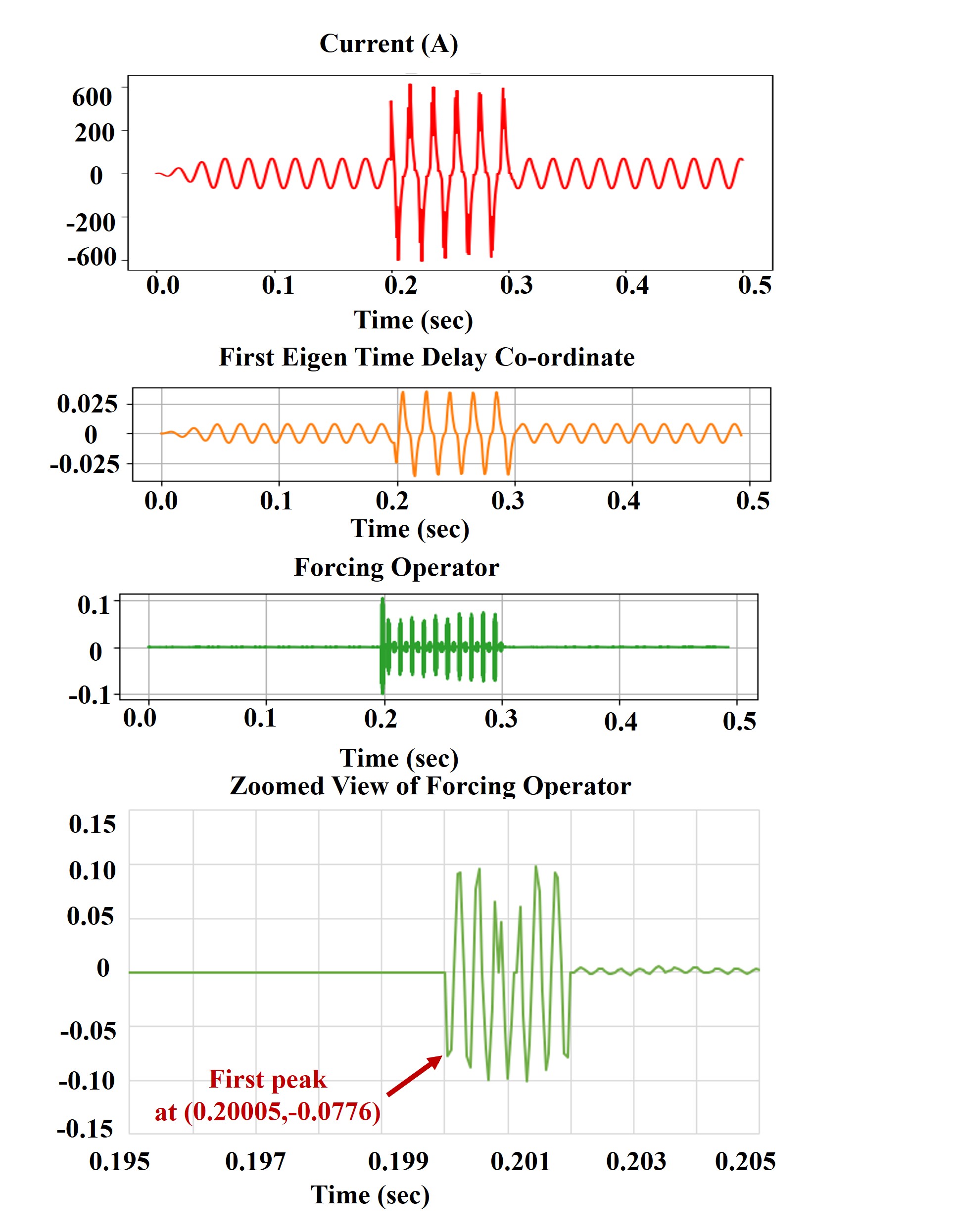}
    \caption{High current Arcing fault with $R_T = 0.001\,\Omega$.}
    \label{fig:FIGno}
  \end{minipage}
\end{figure*}

In this scenario, the \(R_{T}\) value is increased. Low-current arcing faults generally arise in overhead lines, and for this fault class, \( R_T \) is non-zero. A designated current range for low-current arcing faults is employed, along with the correlation between arc voltage and current as outlined in \cite{terzija2004modeling}. Our simulation produces many low-current arcing faults, with parameter values over the whole designated range. Upon executing a simulation for 0.5 seconds with the introduction of an arc fault at 0.2 seconds, the resultant forcing operator deviation is noticed at 0.20045 seconds. This suggests that the system could identify the arc fault within a surprisingly short interval of 0.45 milliseconds. The value of the forcing operator in this particular case is \(-0.1035\), suggesting a very low magnitude and falling within the threshold range for arc faults. This case is depicted in Figure\ref{fig:FIG2}.

\subsection{Case B: High Current Arcing Fault Present} 
High-current arcing faults in MV distribution systems generally arise in underground cables that have low fault resistance. For these fault types, the \( R_T \) value is reduced to its minimum. Our simulation produces an extensive range of high-current arcing faults, guaranteeing that the parameter values encompass the complete spectrum \cite{gaudreau2007evaluation}.
 In this circumstance, the \(R_{T}\) value is incredibly small. Upon conducting the simulation for 0.5 seconds and adding an arc fault at 0.2 seconds, the detection of the forcing operator deviation happens at 0.20005 seconds, confirming the system's capabilities to recognize the arc fault within a quick  0.05 milliseconds. The observed value of the forcing operator in this particular case is -0.0776675
 reflecting a low value. The scenario is shown in Figure \ref{fig:FIGno}.

\subsection{Case C: Arcing Fault Present when grounded in wet cement} 
\begin{figure*}[htp]
  \centering
  \begin{minipage}[b]{0.45\textwidth}
    \includegraphics[width=\textwidth]{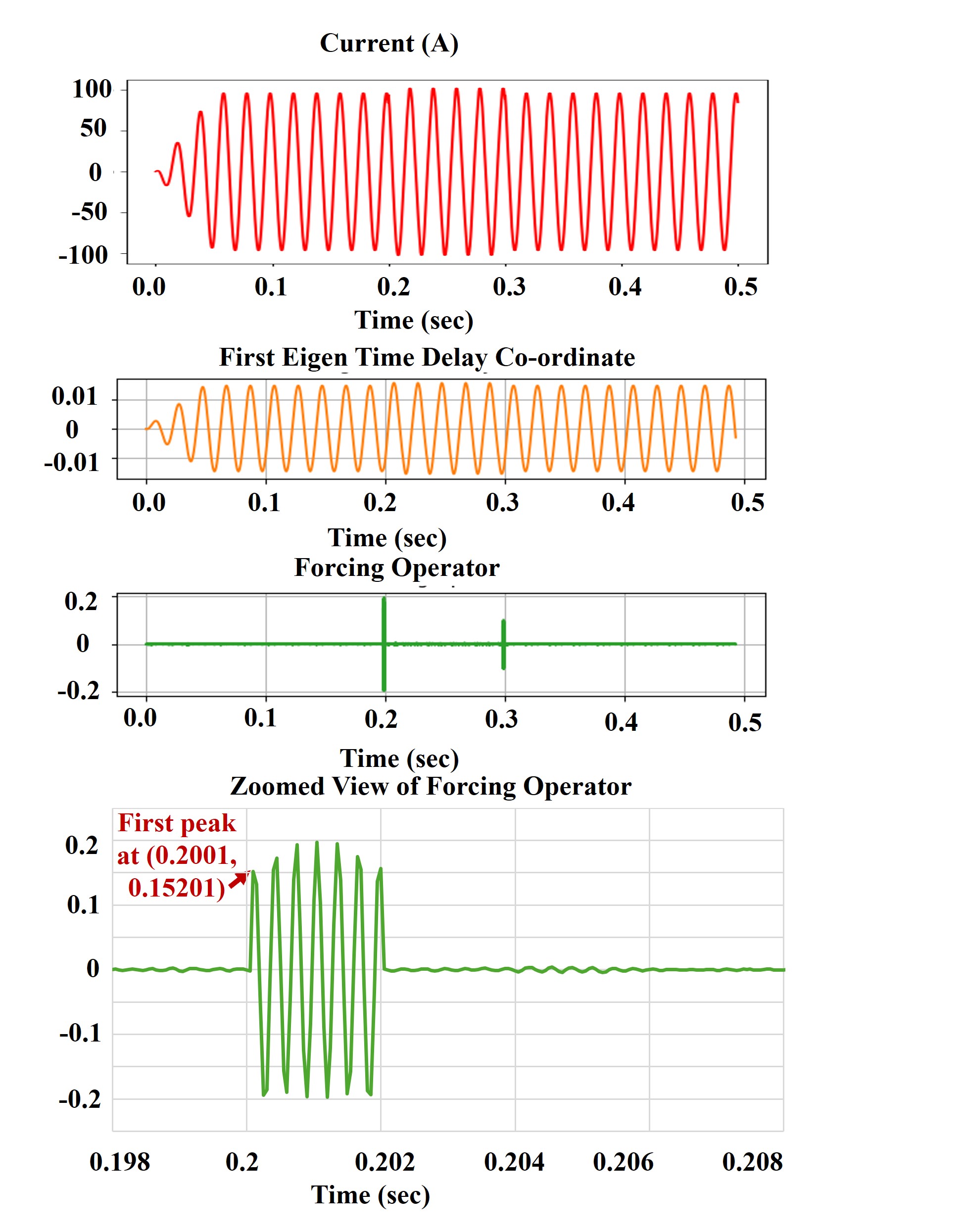}
    \caption{Arcing Fault Present when grounded in wet cement.}
    \label{fig:WETCEM}
  \end{minipage}
  \hfill
  \begin{minipage}[b]{0.45\textwidth}
    \includegraphics[width=\textwidth]{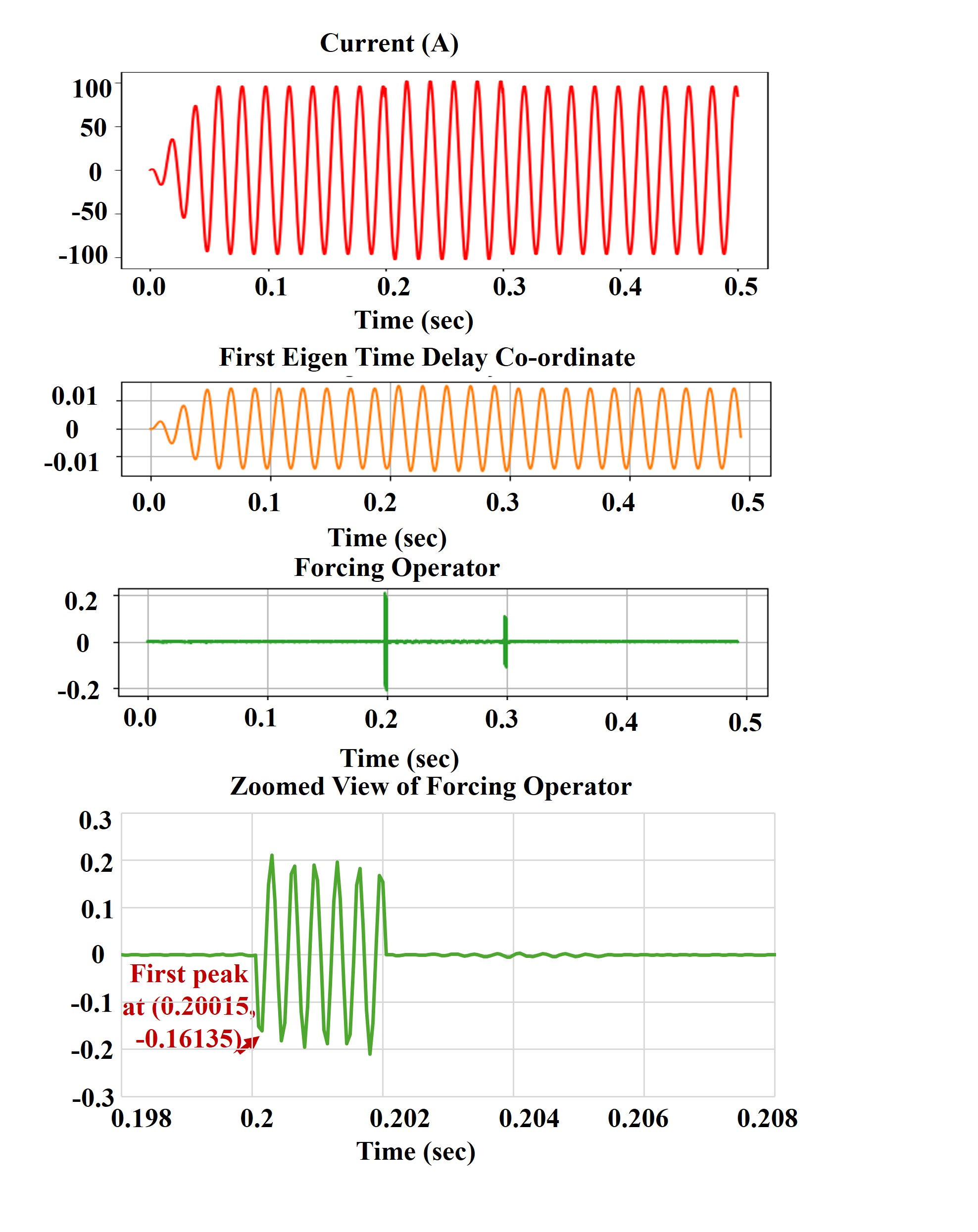}
    \caption{Arcing Fault Present when grounded in dry soil.}
    \label{fig:dr}
  \end{minipage}
\end{figure*}

In practical HIAFs, fault current distortions are greatly affected by multiple parameters, especially surface humidity conditions \cite{ghaderi2017high}. This study investigates HIAF under conditions when it is grounded on wet cement,  with DURATION set to 0.007 seconds and Extent set to 50000 $\Omega$, OFFSET set to 0.2 kV. A simulation was performed for 0.5 seconds, with the arc fault initiated at 0.2 seconds. The subsequent deviation of the forcing operator was recognized at 0.2001 seconds, proving the system's capabilities to identify the arc fault within a short interval of 0.1 milliseconds. The value of the forcing operator in this case was calculated to be 0.15201, which lies inside the threshold range for arc faults. This scenario is represented in Figure \ref{fig:WETCEM}.
\subsection{Case D: Arcing Fault Present when grounded in Dry Soil} 
This study examines HIAF when it is grounded on dry soil. In this study, the DURATION is set to 0.007 seconds, the Extent is defined as 4708 $\Omega$, and the OFFSET is set to 0.2 kV. Simulation was conducted for 0.5 seconds, with the arc fault triggered at 0.2 seconds. The deviation of the forcing operator was identified at 0.20015 seconds, indicating the system's ability to identify the arc fault within a short interval of 0.15 milliseconds. The forcing operator's value in this instance was found to be -0.16135, which falls within the defined threshold range for arc faults. This scenario is represented in Figure \ref{fig:dr}.

\subsection{Case E: Nonarcing disturbance (Load Switching) } 
    
\begin{figure*}[htp]
  \centering
  \begin{minipage}[b]{0.45\textwidth}
    \includegraphics[width=\textwidth]{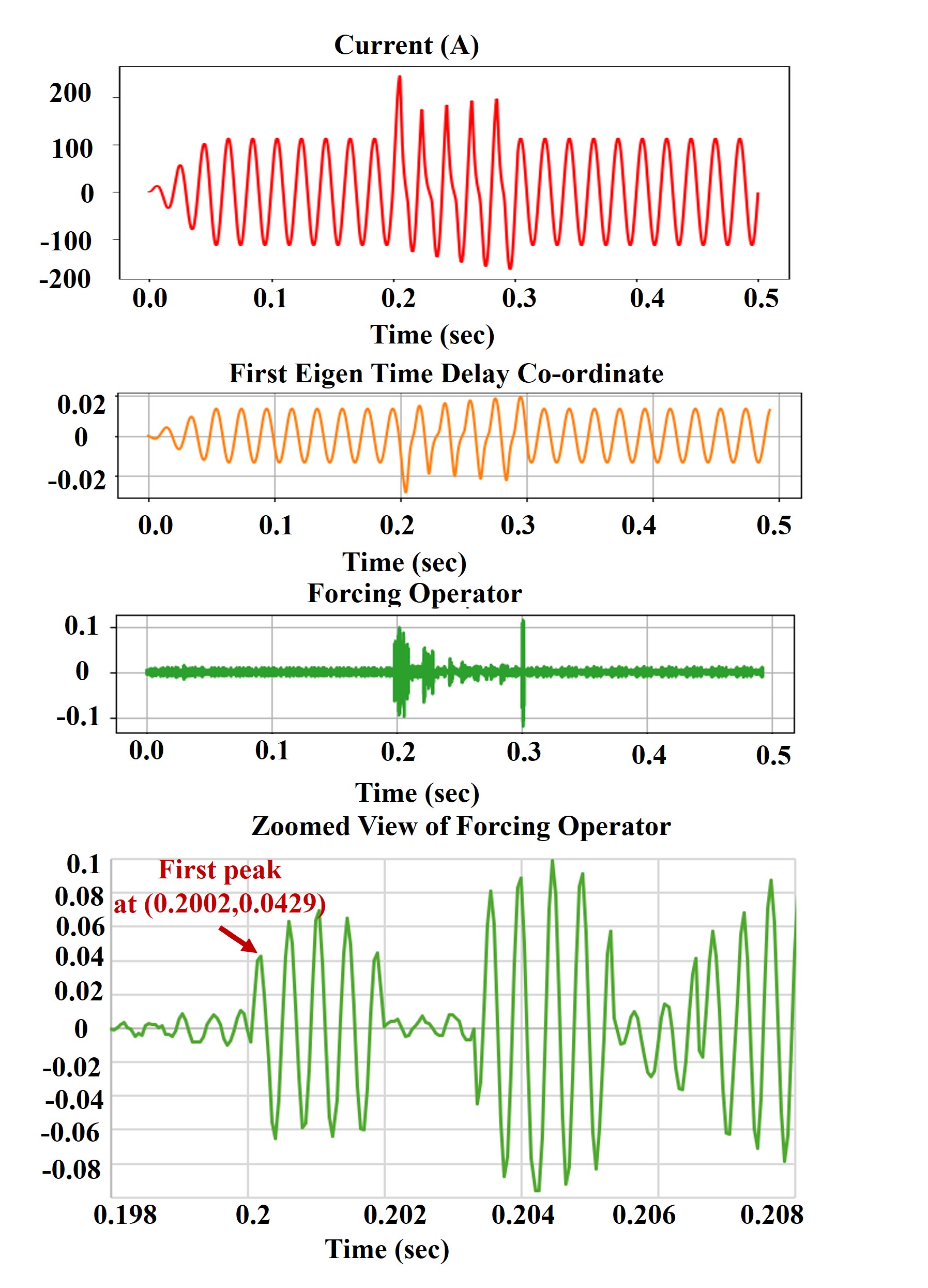}
    \caption{Non-arcing disturbance (load switching).}
    \label{fig:FIGswitch}
  \end{minipage}
  \hfill
  \begin{minipage}[b]{0.45\textwidth}
    \includegraphics[width=\textwidth]{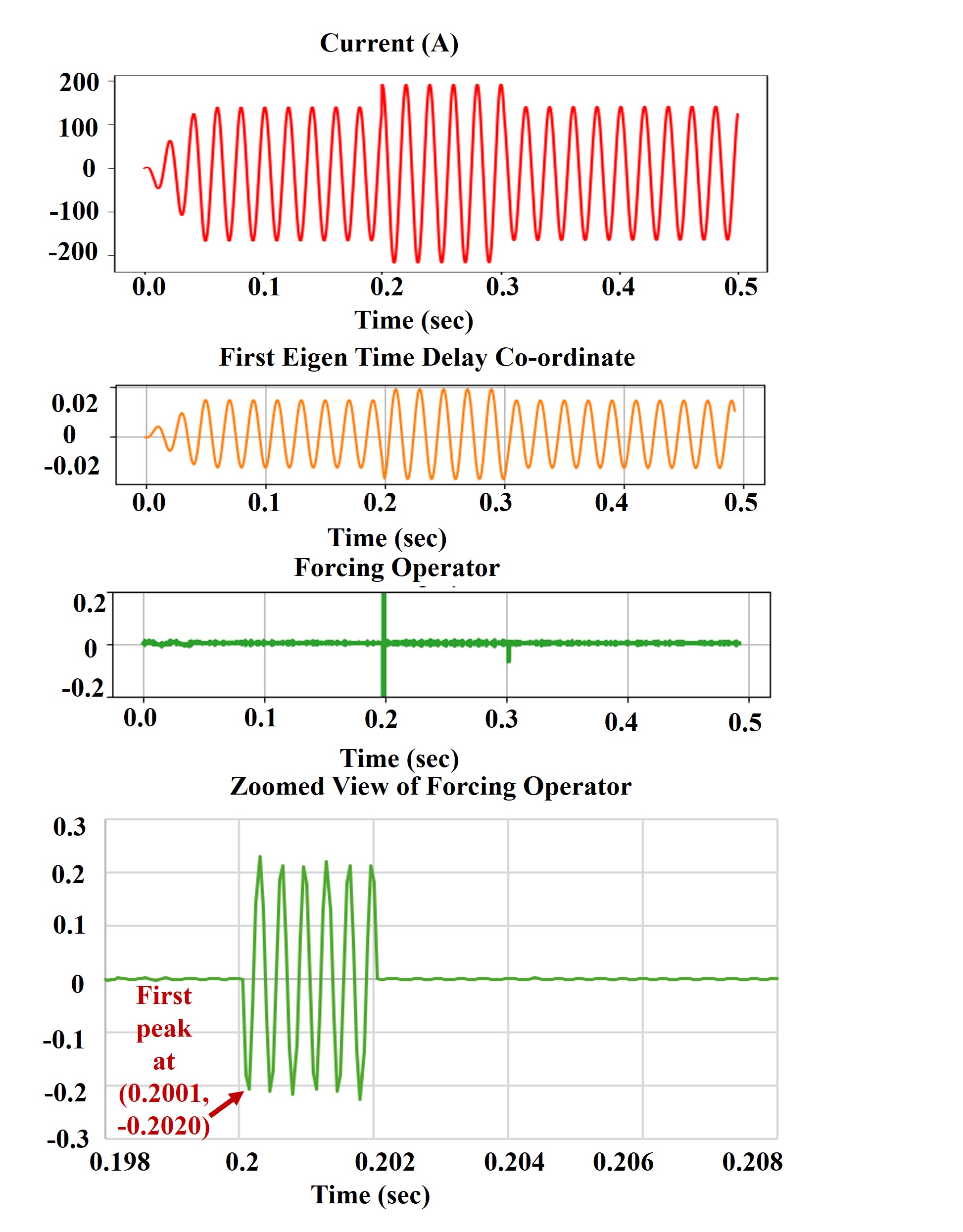}
    \caption{LG (Line-to-ground) fault without arc.}
    \label{fig:FIG5}
  \end{minipage}
  \end{figure*}

 Load switching was employed to distinguish arc faults from non-arcing disturbances. The load was altered between 0.2 and 0.3 seconds, with changes in both location and load type. A variation in the forcing operator was noted at 0.2002 seconds.{ The forcing operator demonstrated a value of \(0.04289\), which falls within the range indicative of non-arcing disturbances.} This situation is depicted in Figure\ref{fig:FIGswitch}.

 \subsection{Case F:  LG (Line-to-Ground) fault without arc.} 
 In this instance, a line-to-ground fault occurs in the system, different from the arc fault scenario. The simulation runs for 0.5 seconds, with the fault introduced at 0.2 seconds and continuing for 0.1 seconds. As a result, the deviation of the forcing operator is seen at 0.2001 seconds, showing the system's ability to identify the issue within 0.1 milliseconds. The value of the forcing operator in this example is measured at -0.202. This case is shown in Figure\ref{fig:FIG5}.

\subsection{Case G: Arcing fault current with Induction Motor Load} 
\begin{figure*}[htp]
  \centering
  \begin{minipage}[b]{0.45\textwidth}
    \includegraphics[width=\textwidth]{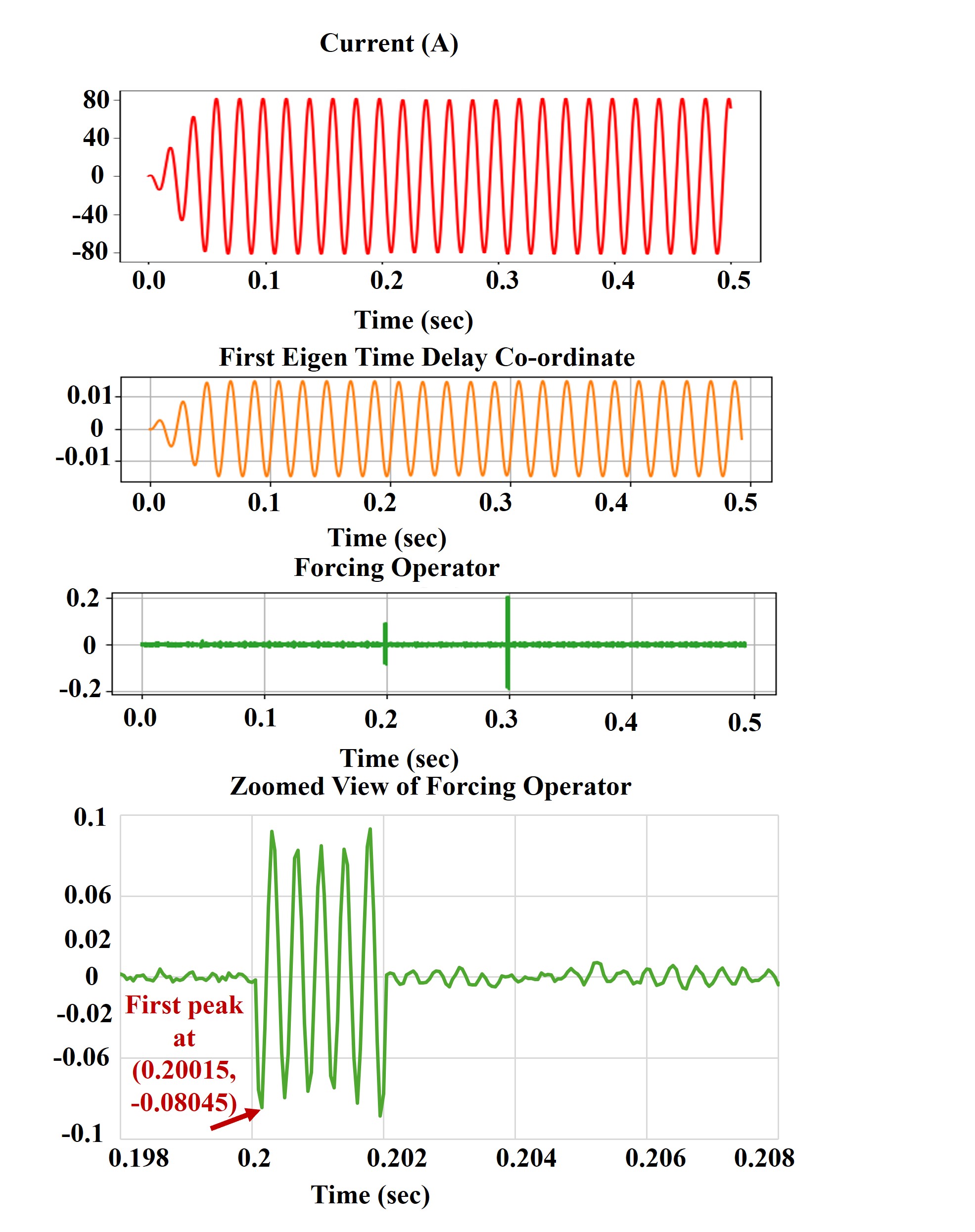}
    \caption{Low current Arcing fault with $R_T = 1000\,\Omega$ with IM load.}
    \label{fig:FIGIMload}
  \end{minipage}
  \hfill
  \begin{minipage}[b]{0.45\textwidth}
    \includegraphics[width=\textwidth]{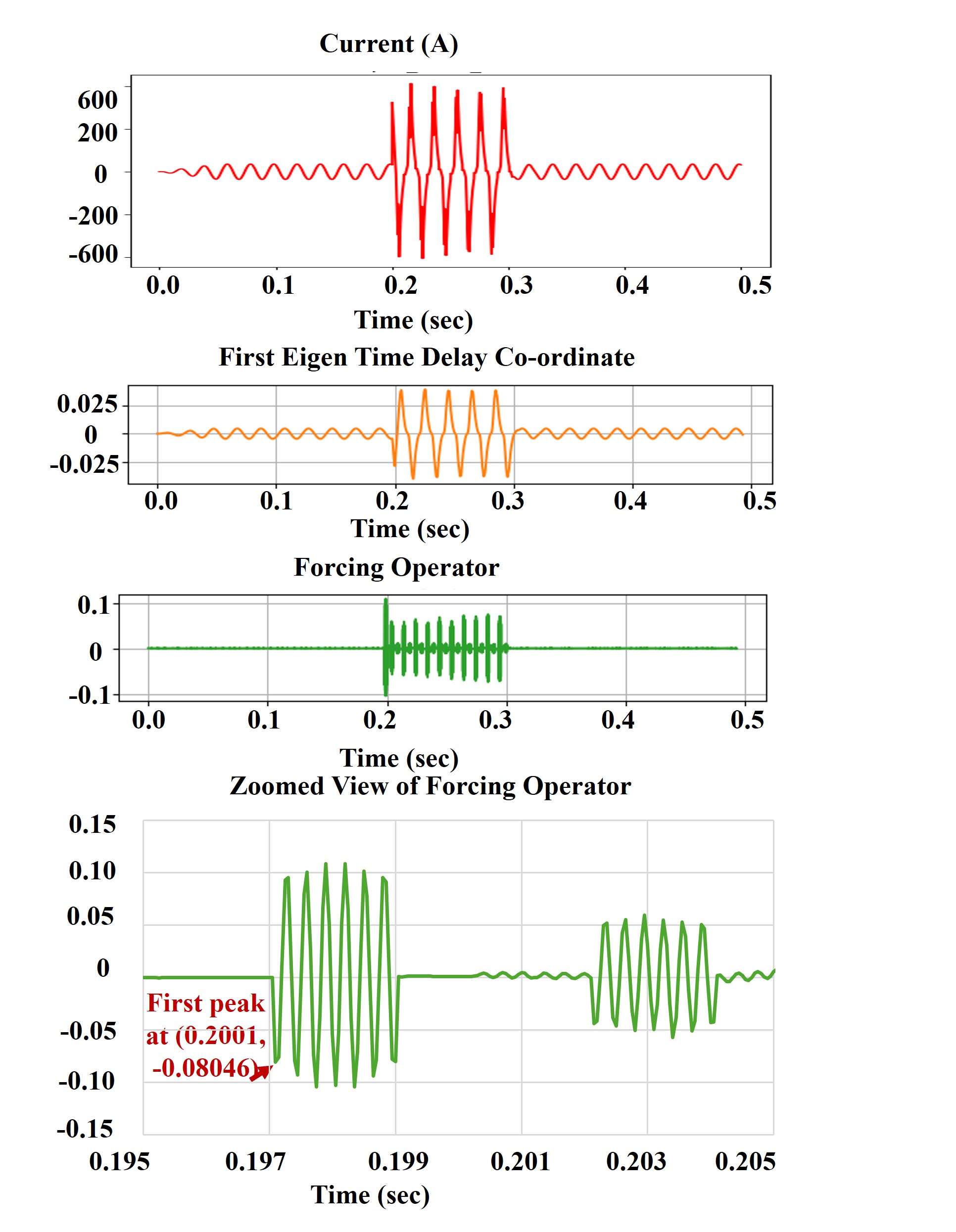}
    \caption{High current Arcing fault with $R_T = 0.001\,\Omega$ with IM load.}
    \label{fig:HCAFIMload}
  \end{minipage}
\end{figure*}

 In this scenario, Load 1 stays constant at 10 MW, while Load 2 comprises an induction motor (IM) operating at 3.3 kV with a rated capacity of 0.2 MVA. The motor is driven through a step-down transformer, which reduces the voltage from 12 kV to 3.3 kV. The simulation runs for 0.5 seconds, with the fault introduced at 0.2 seconds and continuing for 0.1 seconds. Two scenarios have been investigated involving an IM load: one with a low-current arcing fault and the other with a high-current arcing fault. These scenarios are represented in Figures \ref{fig:FIGIMload} and \ref{fig:HCAFIMload}, respectively. In the low-current arcing fault scenario, a deviation in the forcing operator was noted at 0.20015 seconds, with the value of the forcing operator measured to be -0.08045. In the high-current arcing fault scenario, the deviation is visible at 0.2001 seconds, with the value of the forcing operator -0.08046. Both values fall within the threshold range for arc faults, confirming the robustness and effectiveness of our approach. 

\subsection{Case H: Impact of Noise}
\begin{figure*}[htp]
  \centering
  \begin{minipage}[b]{0.6\textwidth}
    \includegraphics[width=\textwidth]{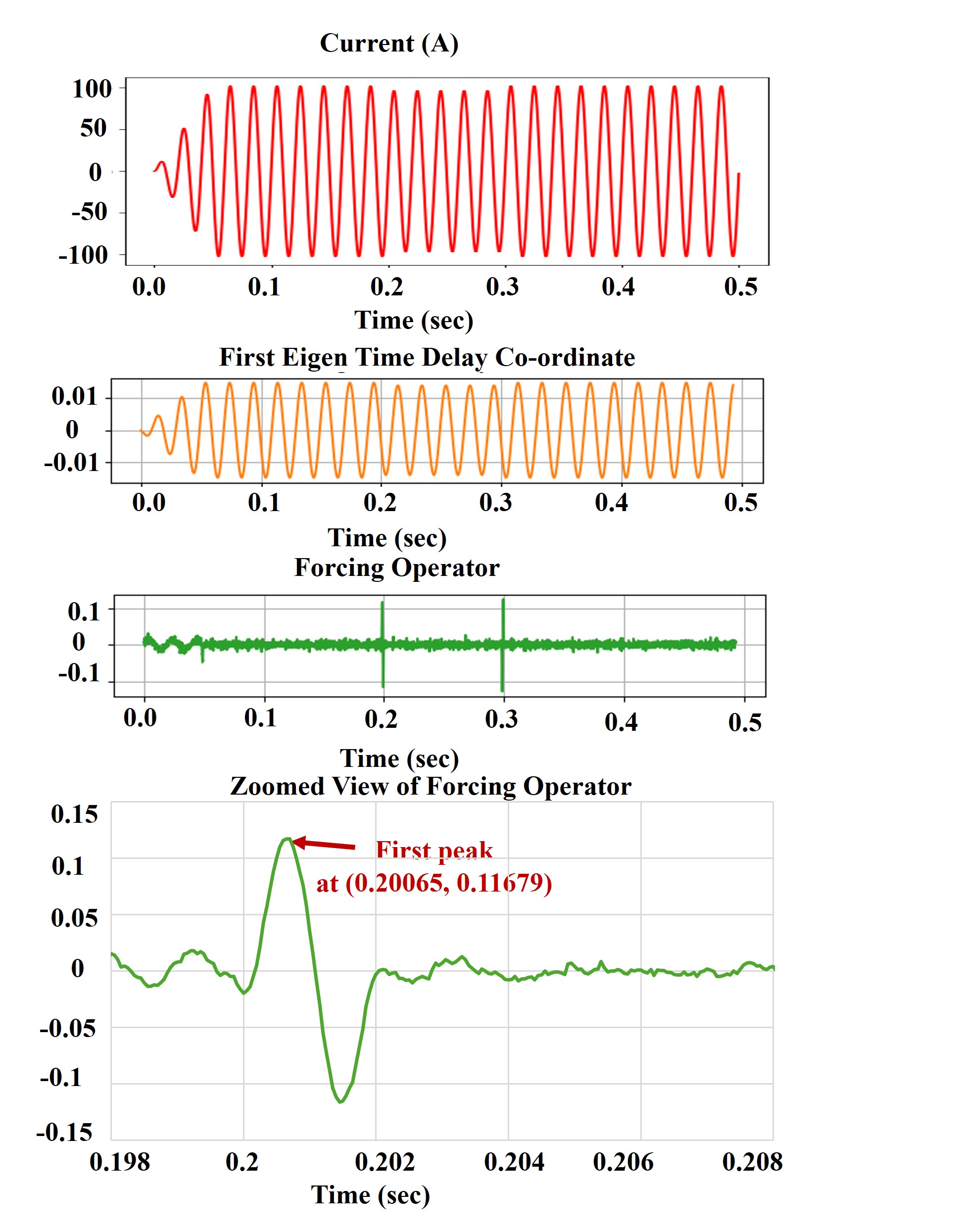}
    \caption{Low current Arcing fault with $R_T = 1000\,\Omega$ with noise.}
    \label{fig:noise}
  \end{minipage}
  \hfill
  \begin{minipage}[b]{0.45\textwidth}
     \includegraphics[width=\textwidth]{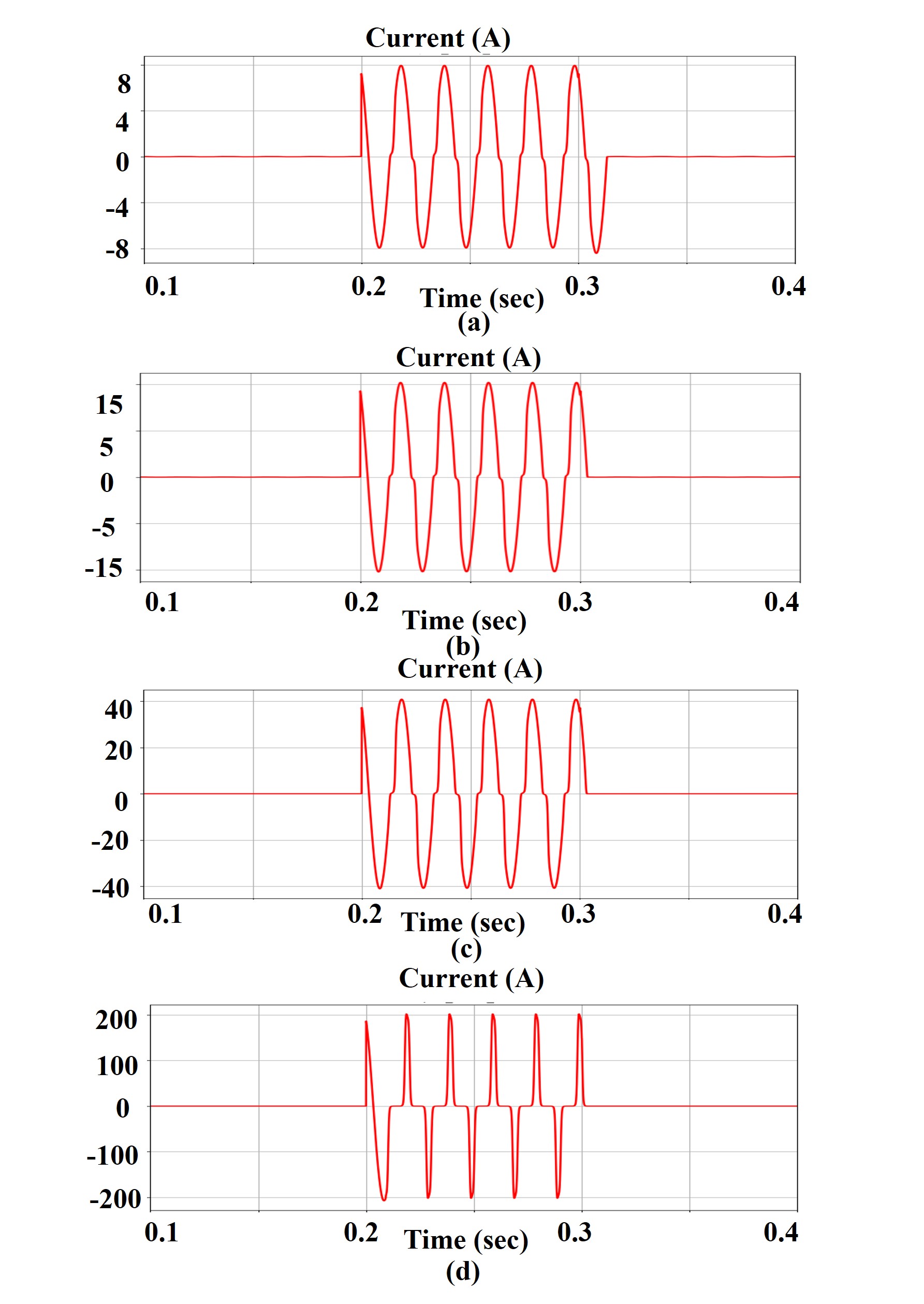}
     
    \caption{Arcing fault current $i_f$  waveforms for varying resistance values: (a) $R_T = 1000\,\Omega$ ( high resistance), (b) $R_T = 500\,\Omega$ (moderate resistance), (c) $R_T = 200\,\Omega$ (lower resistance), and (d) $R_T = 50\,\Omega$ (wet soil, low resistance).
}
    \label{fig:FIG3}
  \end{minipage}
\end{figure*}

 To evaluate the influence of measurement noise, white Gaussian noise is introduced to the current measurements. Simulations were executed with signal-to-noise ratios (SNR) of 60 dB, 70 dB, and above, during which our approach exhibited excellent performance. Incorporating white Gaussian noise at this level and higher guarantees a realistic simulation of noise that distorts the signal without substantially compromising its integrity, thus allowing an appropriate evaluation of the method's effectiveness under realistic circumstances. In cases of extremely low signal-to-noise ratio, denoising filters may be employed for efficient noise suppression. Figure \ref{fig:noise} depicts a scenario with a signal-to-noise ratio of 70 dB. The deviation of the forcing operator is noted at 0.20065 seconds, with a value of 0.116793459. In the presence of noise, it has been found that the detection time is extended. However, the detection is still robust from the performance of our analysis.

 Figure \ref{fig:FIG3}  displays multiple waveforms of fault current \( i_f \) for an arcing fault. By modifying the characteristics of the current—such as its extent, duration, and offset—various arc fault scenarios can be simulated, including conditions like wet soil, low-resistance earthed neutral when DURATION is set to 0.009 seconds and Extent is set to 500000 $\Omega$ OFFSET is set to 0.2 kV, dry soil, isolated neutral when DURATION is set to 0.007 seconds and Extent is set to 4708 $\Omega$ OFFSET is set to 0.2 kV, wet cement, low-resistance earthed neutral when DURATION is set to 0.007 seconds and Extent is set to 50000 $\Omega$ OFFSET is set to 0.2 kV.

\begin{table*}[htbp]
\centering
\renewcommand{\arraystretch}{1}
\caption{Case Study: Detection Time for Different Cases}
\resizebox{\textwidth}{!}{
\begin{tabular}{|c|c|c|c|c|c|c|c|}
\hline
\textbf{Cases} & \textbf{Extent ($\Omega$)} & \textbf{Duration (seconds)} & \textbf{Offset (kV)} & \textbf{$R_T$ ($\Omega$)} & \textbf{Forcing Operator} & \textbf{Time of Detection (ms)}& \textbf{Arc/Disturbance Type} \\
\hline
A & 5000 & 0.00413 & 0.2 & 1000 & -0.1035 & 0.45 & LCAF\\
B & 5000 & 0.00413 & 0.2 & 0.001 & -0.0776 & 0.05 & HCAF \\
C & 50000 & 0.00700 & 0.2 & 50 & 0.1520 & 0.10 & ACWC \\
D & 4708 & 0.00700 & 0.2 & 50 & -0.1614 & 0.15& ACDS \\
E & NA & NA & NA & NA & 0.0429 & 0.20 & NAD (LS)\\
F & NA & NA & NA & NA & -0.2020 & 0.10 & LGF\\
G (a) & 5000 & 0.00413 & 0.2 & 1000 & -0.0805 & 0.15 & LCAFIM\\
G (b) & 5000 & 0.00413 & 0.2 & 0.001 & -0.0805 & 0.10 & HCAFIM\\
H & 5000 & 0.00413 & 0.2 & 1000 & 0.1168 & 0.65 & AFWN\\
\hline
\end{tabular}
}

\label{tab:compare_dataset}

\end{table*}
\vspace{0.3em}

\noindent \textbf{Notation:}
LCAF — Low Current Arc Fault, HCAF — High Current Arc Fault, NAD (LS) — Non-Arcing Disturbance (Load Switching), LGF — Line to ground fault , LCAFIM — Low Current Arc Fault with Induction motor load, HCAFIM — High Current Arc Fault with Induction motor load, AFWN — Arc Fault with Noise.

Table \ref{tab:compare_dataset} provides an overview of the analysis of the experiments, including the corresponding time of detection for each conducted experiment.
The range of the forcing operator permits the differentiation between arc faults and other types of faults, as arc faults present a lower range and require more time for detection compared to other fault types. It has been shown that a forcing operator range of  +0.06 to + 0.18 and -0.06 to -0.18 helps the rapid identification of arc faults. In contrast, when the forcing operator range changes, such as +0.2 and higher values and -0.2 and lower values, the system identify other faults, while non-arcing disturbances, such as load switching, are identified with a forcing operator range of +0.045 to -0.045.To assess detection robustness, we conducted 50 simulations for each scenario. The proposed technique attained a detection accuracy of 100\% for arc faults, with 0\% false positives for non-arcing disturbances. The detection exhibited consistency across the trials, proving its robustness.

\section{Comparison With Existing Techniques}
\label{section:Comparison With Existing Methods}
A comparative analysis has been conducted between the proposed method and existing arcing fault detection method found in the literature. The standard detection time for high-impedance arc faults in medium voltage networks may vary based on the specific detection methods and technologies employed. Generally, high-impedance arc faults are hard to detect because of their low fault current and stochastic nature, which can result in longer detection times compared to other fault types. Usually, detection times for high-impedance faults vary between milliseconds to several seconds.Our proposed method is carefully assessed against the method outlined in \cite{8798732}, employing the same dataset for a direct and thorough comparative study. This method in \cite{8798732} identifies arc faults by examining waveform features through two approaches: it employs the Fast Fourier Transform (FFT) to measure harmonic unpredictability, signifying unstable arcing, and the Discrete Wavelet Transform (DWT) to evaluate waveform distortion, indicating stable arcing. The properties are assessed against thresholds established from a healthy signal baseline to ascertain the occurrence and duration of arc faults.The technique detects HIAFs in 70 milliseconds. Improvements in technology continue to push for faster and more reliable detection systems in medium voltage networks. However, our proposed method demonstrates the early detection capability of arc faults. The findings, as outlined in Table \ref{tab:Comparison}, indicate that the suggested method attains a detection time of 0.45 ms, which is  99.36\% enhanced than the Harmonic Randomness and Waveform Distortion approach (70 ms) and exceeds the performance of other  techniques by greater margins.  

The findings demonstrate that the suggested method is both efficient and resilient under diverse demanding settings, including low and high current arc faults, different grounding materials, and noisy environments.  The method's dependence on Koopman-invariant forcing coordinates enables it to accurately differentiate arc faults from non-arcing disturbances, a task that many black-box AI methodologies find challenging.  The method's reliable performance, even amongst measurement noise (with an SNR as low as 60 dB), underscores its practical feasibility.

 Moreover, the physics-informed foundation of the HAVOK model offers interpretable features, which is a significant benefit in safety-critical fields where explainability is as vital as precision.  This study establishes a solid basis for future arc fault detection systems in intricate electrical networks by the integration of rapid detection, resilience, and interpretability.

\begin{table}[h]
\centering
\caption{Comparative Summary of Existing Arc Fault Detection Techniques and Advancements by the Proposed Method}
\label{tab:comparison}
\resizebox{\textwidth}{!}{%
\begin{tabular}{|p{3cm}|p{3.5cm}|p{2.5cm}|p{4.5cm}|p{4.5cm}|}
\hline
\textbf{Method} & \textbf{Technique} & \textbf{Detection Time} & \textbf{Limitations} & \textbf{Advancement by This Work} \\
\hline
Harmonic Randomness and Waveform Distortion\cite{8798732} & Harmonic analysis of current signals & 70 ms & Conventional harmonic and DWT techniques may exhibit sensitivity to fluctuating load circumstances, noise, and non-arc disturbances, which could result in false positives or undetected anomalies without careful calibration and a solid baseline foundation.& 99.36\% enhaced detection \\
\hline
\textbf{Proposed Method} & Koopman operator-based HAVOK with forcing signature & \textbf{0.45 ms} & Requires parameter tuning for unseen networks & Fast, explainable detection using Koopman-invariant forcing operator; resilient under noise and non-ideal conditions \\
\hline
\end{tabular}%
}

\label{tab:Comparison}
\end{table}

\section{Conclusion}
\label{section:conclusion}
Our research greatly contributes to the field of arc fault detection by addressing crucial gaps in existing knowledge. The application of data obtained by PSCAD simulations of medium voltage distribution lines distinguishes our analysis, filling a noteworthy gap in earlier research largely focused on low-voltage systems. This study proposed a unique, physics-informed methodology for detecting high-impedance arc faults (HIAFs) in MV electrical distribution systems using the HAVOK analysis.  By integrating Koopman operator theory into nonlinear current dynamics, the method made major improvements in detection speed and interpretability over traditional and black-box AI-based techniques.  The suggested approach successfully detects arc faults under a wide range of situations, including different grounding surfaces, noise levels, and load types, achieving detection within as little as 0.45 milliseconds.  Comparative investigation indicated that the strategy is 99.36\% enhanced over other existing methods, while maintaining good detection accuracy.  These contributions boost the reliability and responsiveness of arc fault detection systems, representing a considerable advancement in the field of power system safety.

We acknowledge that the present study is confined to simulation-based validation.  Field testing of medium-voltage systems presents practical and safety concerns; future efforts will incorporate hardware-in-the-loop (HIL) configurations and experimental validation with lab-scale prototypes to further substantiate the suggested detection method. For future work, the proposed approach should be tested using real-world datasets obtained from physical testbeds or operational distribution networks to ensure its robustness outside of simulation. The insights provided here pave the path for future developments in the domain, promoting a more secure and trustworthy operational environment.

\bibliography{main}

\end{document}